\documentstyle[12pt,epsf]{article}
\newcommand{\beq}{\begin{equation}}
\newcommand{\eeq}{\end{equation}}
\newcommand{\beqa}{\begin{eqnarray}}
\newcommand{\eeqa}{\end{eqnarray}}
\def\nue{{\nu_e}}
\def\anue{{\bar{\nu_e}}}

\newcommand{\dm}{\mbox{$\Delta{m}^{2}$~}}

\begin{document}


\begin{center}
{\large{ \bf { Status of the neutrino decay solution to the solar 
neutrino problem}}}
\vskip 10pt
{\it  Sandhya Choubey$^a$\renewcommand{\thefootnote}{\fnsymbol{footnote}}
\footnote{sandhya@tnp.saha.ernet.in}, 
Srubabati Goswami$^a$\renewcommand{\thefootnote}{\fnsymbol{footnote}}
\footnote{sruba@tnp.saha.ernet.in},
Debasish Majumdar$^b$\renewcommand{\thefootnote}{\fnsymbol{footnote}}
\footnote{debasish@tnp.saha.ernet.in\\
$~~~~~^b$Present Address: Saha Institute of Nuclear Physics, Calcutta, INDIA.}\\
$^a$Saha Institute of Nuclear Physics,\\1/AF, Bidhannagar,
Calcutta 700 064, INDIA.\\
$^b$92 Acharya Prafulla Chandra Road, Calcutta 700 009, INDIA.\\}
\end{center}
\vskip 30pt

\begin{center}
Abstract
\end{center}

We re-examine the neutrino decay solution to the solar neutrino
problem in the light of the SuperKamiokande (SK) data. 
For the decay solution 
the SK spectrum data by its own can provide a fit comparable to
the fit obtained from the MSW solution. However 
when one combines the results from the total rates 
of the 
$^{37}{Cl}$ and $^{71}{Ga}$ experiments the fit becomes much poorer. 

keywords: neutrino mass, solar neutrinos, decay, mixing

\newpage

In this paper we examine the status of the neutrino decay
solution to the solar neutrino problem in the light of the 
SK data\renewcommand{\thefootnote}{\arabic{footnote}}
\setcounter{footnote}{0}\footnote{The possibility of neutrino 
decay as a solution to the
atmospheric neutrino problem has been considered recently in \cite{atmdk,sc}}. 
Neutrino decay as a 
solution to the solar neutrino problem has been considered earlier (pre-SK)
\cite{app,ajp,acker}.
However since the Ga data implies that the low energy pp neutrinos are 
less suppressed compared to the high energy $^{8}{B}$ suppression seen in 
Cl or the Kamiokande experiments, this solution was ruled out at 99\% 
C.L. since the decay term $\exp(-\alpha L/E)$, (where $\alpha=m_2/\tau_o$, 
$m_2$ being the mass of the unstable state and $\tau_o$ its rest frame
lifetime), 
suppresses the low energy neutrino flux more than the high energy flux. 
However the SK spectrum data shows more events for the high energy 
bins.  Though the statistics in these high energy bins need improvement, we 
explore the status of the neutrino decay 
scenario in the context of this behaviour of the SK spectrum data.

Radiative decays of neutrinos are severely constrained \cite{fukugita} and
thus we are interested in the non-radiative decay modes.
Two classes of models have been considered in the literature in this connection.
\begin{enumerate}
\item In \cite{app} neutrinos are considered to be Dirac particles. 
They consider the decay mode $\nu_2 \rightarrow \bar{\nu}_{1R} + \phi$, 
where ${\bar{\nu}}_{1R}$ is a right handed singlet and $\phi$ is an
iso-singlet scalar. Thus all the final state particles are sterile. 

\item In \cite{ajp} neutrinos are assumed to be Majorana particles and 
the decay mode is $\nu_2 \rightarrow \bar{\nu}_1 + J$, where $\bar{\nu}_1$
interacts as a $\bar{\nu}_e$ with a probability $|U_{e1}|^2$ and 
J is a Majoron. 
\end{enumerate}

We work with just two flavors for simplicity and assume 
that the state $\nu_2$ is unstable, which decays  
with a rest frame lifetime $\tau_o$. 
The other neutrino mass states have lifetimes much greater than the 
sun-earth transit time and hence can be taken as stable. 
In presence of 
decay,
\begin{eqnarray}
P_{\nue\nue} &=& (1-|U_{e2}|^2)^2+|U_{e2}|^4\exp(-4\pi L/\lambda_d)\nonumber 
\\
&&{} + 2 |U_{e2}|^2 (1 - |U_{e2}|^2) \exp(-2 \pi L/\lambda_d)
\cos(2 \pi L/\lambda_{osc})
\label{peedo} 
\end{eqnarray}
\begin{eqnarray}
P_{\nue\nu_x} &=& |U_{e2}|^2 (1-|U_{e2}|^2)\{1+\exp(-4\pi L/\lambda_d)
\nonumber
\\
&&{} - 2 \exp(-2 \pi L/\lambda_d) \cos(2 \pi L/\lambda_{osc})\}
\label{pexdo}
\end{eqnarray}
where $x$ can be either $\mu$ or $\tau$ in the two flavour framework. 
$\lambda_d$ is the decay length defined as,
\begin{equation}
\lambda_d = 2.5\times 10^{-3} km \frac{E}{MeV} \frac{eV^2}{\alpha}
\end{equation}
$\lambda_{osc}$ is the oscillation wavelength defined as 
\begin{equation}
\lambda_{osc} = 2.5\times 10^{-3} km \frac{E}{MeV} \frac{eV^2}{\Delta m^2}
\end{equation}
$L=R(t)(1-r/R(t))$, r being the distance of the point of 
neutrino production from the center of the sun and $R(t)$ is the 
sun-earth distance given by,
\begin{equation}
R(t)=R_0\left[1-\epsilon\cos(2\pi\frac{t}{T})\right]
\end{equation}
Here, $R_0=1.49\times 10^{13}$ cm is the mean sun-earth distance 
and $\epsilon=0.0167$ is the ellipticity of the earth's orbit, $t$ is the 
time of the year and $T$ is 1 year. 
The $\Delta m^2$ dependent oscillations are important around $10^{-10} -
10^{-11}$ eV$^2$. We assume $\Delta m^2$ to be much higher 
so that the cosine term averages out to
zero. We further assume $U_{e1}= \cos \theta$ and $U_{e2} = \sin 
\theta$ so that the survival probability is   
\begin{equation}
P_{\nue\nue} = \cos^4 \theta + \sin^4\theta \exp(-4\pi L/\lambda_d)  
\label{pee}
\end{equation}

In Fig. 1 we plot $P_{\nue \nue}$ as a function 
of $\alpha$ for $\sin^2\theta=0.51$ 
for two illustrative values of $\nue$ energies 7 and 13 MeV. 
For $\alpha < 10^{-13} eV^2$, the $\exp(-4\pi L/\lambda_d) 
\approx 1$ signifying very little decay. Beyond  $\alpha = 10^{-10} eV^2$ 
the $\exp(-4\pi L/\lambda_d) \approx 0$ which is the case where we have 
very fast decay. In both these regions the probabilities are 
energy independent. In the region where $\alpha $ is around $10^{-11} eV^2$ 
the probability does depend on the energy.

The details of the solar neutrino code employed for performing the
$\chi^2$-analysis is described in \cite{gmr1,gmr2}. 
We use the BP98 solar model \cite{bp98} as the reference SSM.  
We perform $\chi^2$-analyses    
\begin{itemize}
\item using the total rates from $^{37}{Cl}$, $^{71}{Ga}$ and SK
experiments. We incorporate the theory errors and their correlations.  
\item using the 825 day SK spectrum data \cite{825}
 including the uncorrelated as well as the 
energy-bin correlated errors.
\item of the combined rates and SK spectrum data
\end{itemize}
For the last two cases we vary $X_{B}$ -- 
the normalization of the $^{8}{B}$ flux with respect to the SSM value as 
free parameter and determine its best-fit value. 

From the fact that the neutrinos from SN1987A have not decayed on their way 
one gets a lower bound 
on the electron neutrino lifetime as $\tau_{0} > 5.7
\times 10^{5} (m_{\nu_e}$/eV)
sec. However if one includes neutrino mixing then shorter lifetimes are 
allowed provided $|U_{e2}|^2 < 0.81$ \cite{haber}. 
To be consistent with this, in our analysis we vary $\sin^2 \theta$ in
the range 0 to 0.8.
 
The data used for the total rates is presented in Table 1. 
Since SK has much 
higher statistics, we have not included the Kamiokande results.
The 
results obtained for Model 1 are summarised in Table 2.
For the 
Model 2 $\nu_2$ decays to $\bar\nu_1$, which interacts as $\anue$ with 
a probabitity $U_{e1}^2$ and $\bar\nu_x$ with a probability $U_{e2}^2$. 
We present in Table 3 the results for Model 2, where we have taken into 
account the $\anue-e$ and $\bar\nu_x - e$ scattering in addition to the 
$\nue -e$ scattering in SK, while for the radiochemical 
experiments there is no change. Since the $\bar{nu_{1}}$ is degraded
in energy \cite{ajp} the effect is not significant.

\vskip 8mm

\begin{center}
\begin{tabular}{|c|c|c|c|}
\hline
Experiment & Chlorine & Gallium & Super-Kamiokande \\ \hline
$\frac{\rm Observed \;\; Rate}{\rm BP98 \;\; Prediction}$ &
$0.33 \pm 0.029$ & $0.562 \pm 0.043$ & $0.471 \pm 0.015$ \\ \hline
\end{tabular}

\vskip 3mm
\parbox{5in}{
Table 1: The ratio of the observed solar neutrino rates to the
corresponding BP98 SSM predictions used in this analysis. The results
are from refs. \cite {solar} and \cite {825}.  For Gallium, the
weighted average of the SAGE and Gallex results has been used.}
\end{center}

\vskip 8mm

\begin{center}
\begin{tabular}{|c|c|c|c|}
\hline
Parameters & Rates & Spectrum & Rates+Spectrum \\
 & (d.o.f = 1) & (d.o.f = 15) & (d.o.f = 18) \\
\hline 
$\alpha$ (eV$^2$) & 0  & 4.22 $\times 10^{-11}$ & 3.29 $\times 10^{-12}$\\
\hline
$\sin^2 \theta$ &0.5 & 0.51 & 0.32 \\ \hline 
$X_{B}$ & Fixed at SSM value & 1.5 & 0.72  \\
\hline
$\chi^2_{min}$ & 12.59 & 17.68 & 33.59 \\ \hline
g.o.f & 3.88 $\times 10^{-2}$ \% & 27.99\% & 1.41\% \\ 
\hline
\end{tabular}

\vskip 3mm
\parbox{5in}{
Table 2: The best-fit values of the parameters, the $\chi^2_{\min}$
and the goodness of fit (g.o.f) for Model 1.}
\end{center}

\vskip 8mm

\begin{center}
\begin{tabular}{|c|c|c|c|}
\hline
Parameters & Rates & Spectrum & Rates+Spectrum \\
 & (d.o.f = 1) & (d.o.f = 15) & (d.o.f = 18) \\
\hline 
$\alpha$ (eV$^2$) & 0  & 3.3 $\times 10^{-11}$ & 3.29 $\times 10^{-12}$\\
\hline
$\sin^2 \theta$ &0.5 & 0.53 & 0.32 \\ \hline 
$X_{B}$ & Fixed at SSM value & 1.5 & 0.71  \\
\hline
$\chi^2_{min}$ & 11.98 & 17.62 & 33.12 \\ \hline
g.o.f & 5.38 $\times 10^{-2}$ \% & 28.32\% & 1.61\% \\ 
\hline
\end{tabular}

\vskip 3mm
\parbox{5in}{
Table 3: The best-fit values of the parameters, the $\chi^2_{\min}$
and the goodness of fit (g.o.f) for Model 2.}
\end{center}

\vskip 8pt

From the analysis of only rates data the best-fit comes in the region where
From the analysis of only rates data the best-fit comes in the region where
$\alpha = 0$, implying that the neutrinos are stable.  
For the total rates, the low energy neutrino flux should be suppressed 
less which is in contradiction with the energy dependence of the 
exponential decay term. Thus the best fit comes in the region where the 
decay term goes to 1. Since the probability in this region is 
energy independent the quality of the fit is not good.   

For the only spectrum analysis, the best-fit comes in the region 
where the $\exp(-\alpha L /E)$ term is non-vanishing and  
the high energy neutrinos are suppressed less.
Since high energy bins have more number of events, the fit is much
better compared to the fit to the total rates and   
is comparable to the ones obtained for the MSW oscillation 
solution ($\chi^2_{min}= 17.62$) \cite {gmr2}. 
The vacuum oscillation solution gives a better fit \cite{gmr1}.
The best-fit values quoted in Table 2 are obtained 
with $X_{B}$ constrained to be $\leq$ 1.5. 
However if we remove the upper limit 
and allow $X_{B}$ to take arbitrary values  
a slightly better fit is obtained for very high values of $X_{B}$:  
\begin{itemize}
\item $\alpha = 6.53 \times 10^{-11}$ eV$^2$, $\sin^2 \theta =0.8$, 
$X_{B} = 6.42, \chi^2_{min} =  17.22, g.o.f.=30.41\% $
\end{itemize} 
We note that the $\alpha$ does not change much. But at higher 
$X_{B}$ one needs a higher $\theta$. 
To understand these features, in fig. 2 we plot $X_{B}$ vs $\alpha$
for various fixed values of $\sin^2 \theta$.
For getting this curve we determine the $X_{B}$ that gives the 
minimum $\chi^2$ at a particular $\alpha$ and then repeat this 
excersise for $\alpha$ varying in the range $10^{-14} - 10^{-8}$ eV$^2$.
The minimum $\chi^2$ obtained at each point of the parameter space of fig. 2 
is within the 90\% C.L. limit of the global $\chi^2$ minimum.  
There are three regions 
\begin{enumerate}
\item For very low values of $\alpha$ (upto $10^{-13}$ eV$^2$) the exponential
term is 1 and $P_{\nue \nue}$ = 1 - 0.5 $\sin^2 2 \theta$ which is the 
average oscillation probability. This can vary from 1 to 0.5 depending on
$\theta$. 
However in 11 out of 18 bins of the SK spectrum data, 
the rate$_{\rm obs}$/SSM $<$ 0.5 and this can be
achieved only by keeping $X_{B} < 1.0$.  

\item In the range $10^{-13} - 10^{-10}$ eV$^2$ the exponential term
contributes to the survival probability and it falls sharply with increasing 
$\alpha$. Therefore  in this range $X_{B}$ rises sharply
as $\alpha$ increases.

\item Beyond $10^{-10}$ eV$^2$ the exponential term goes to zero and 
the probability is $P_{\nue \nue} = cos^4 \theta$. In this zone
one can achieve a probability $<$ 0.5 by adjusting $\theta$ only
and $X_{B}$ does not play an important role. However if $X_{B}$ is allowed 
to vary then 
for smaller values of $\theta$ the $X_{B}$ needed is low
and vice-versa. 
 
\end{enumerate} 

In fig. 3 we plot the $\chi^2$ for the SK spectrum data, 
as a function of one of the parameters, 
keeping the other two unconstrained.
In fig. 3a the solid(dashed) line gives the variation of 
$\chi^2$ with $\alpha$ keeping $X_{B}$ unconstrained (fixed at 1.0). 
If $X_{B}$ is 1 then lower values of $\alpha$ are not 
allowed as by varying $\theta$ alone one cannot get a probability 
less than 0.5. If $X_{B}$ is allowed to vary then low values of $\alpha$ 
are also admissible. 
Higher values of $\alpha$ are allowed for both the cases. 
The minimum of course comes in the region where $\alpha \sim 10^{-11}$ eV$^2$
for which the high energy neutrino events are less suppressed. 

Fig. 3b gives the variation of $\chi^2$ with $\sin^2 \theta$ keeping  
$X_{B}$ unconstrained (solid line) and $X_{B}=1$ (dashed line). 
For both the curves $\alpha$ can take any value. 
For $X_{B}$ = 1 one cannot get a good fit in the 
low $\alpha$ region as discussed above and the fit goes to the high 
$\alpha$ region. In this zone the probability ($\sim \cos^4 \theta$) 
increases as we decrease $\theta$ and this puts a lower limit on
the allowed value of $\theta$. This can be evaded if $X_{B}$ is allowed
to vary, as by adjusting $X_{B}$ one can get a good fit even if $\theta$ is
very low.   
 
Fig. 3c gives the allowed range of $X_{B}$ keeping the other two 
parameters unconstrained.
From the figure we see that values of $X_{B}$ below 0.4 are not allowed
at 90\% C.L.. As discussed,
$X_{B}$ plays an important role in the low and 
intermediate $\alpha$ regions.
In the latter zone the $X_{B}$ required is high, 
therefore the constraints on the low values of $X_{B}$ comes from the low 
$\alpha$ region. 
In this region as we decrease $X_{B}$ the number of events will decrease
which can be adjusted by increasing the survival probability 
and the maximum value of $P_{\nue\nue}=1.0$  
gives a lower limit on $X_{B}$. 
For very high values of $X_{B}$ the best-fit goes in the 
high $\alpha$ region where the probability is $\sim cos^4\theta$. 
As we increase $X_{B}$ the number of events will increase which 
can be compensated by increasing $\theta$. However from 
SN1987A constraints $\sin^2 \theta$ is restricted to be $<$ 0.8 and 
therefore beyond 
$X_{B} = 7.2$ one does not get a good fit.  

The rest frame lifetime obtained at the best-fit $\alpha$ 
from the spectrum data is $\tau_o = (m_2/\rm{eV}) 
1.54 \times 10^{-5}$ sec. This is small enough 
for decay to happen before neutrino decoupling in the early universe. This 
can increase the effective number of light neutrinos, $N_{\nu}$,
from 3. However depending on the data used the upper limit on
$N_{\nu}$ can be 5 or 6 \cite{cosmo} which is consistent with the model of
neutrino decay used here. 

The rest frame lifetime of $\nu_{2}$ is given by \cite{atmdk}
\begin{equation}
\tau_{0} = \frac{16 \pi}{g^2} \frac{m_{2} (1 + m_{1}/m_{2})^{-2}}{\dm}
\end{equation}
From the best-fit $\alpha$ $\sim 10^{-11}$ eV$^2$ and assuming 
$m_{2} >> m_{1}$ one gets 
\begin{equation}
g^2 \dm \sim 16 \pi 10^{-11}
\end{equation} 
If we now incorporate the bound $g^2 < 4.5 \times 10^{-5}$ as obtained
from K decay modes \cite{barger} we obtain  
$\dm {>}{\sim}10^{-5}$ eV$^2$, 
consistent with our assumption.
At the best-fit $\alpha$ the decay length for atmospheric neutrinos 
is $\lambda \approx 2.5 \times 10^{11}$ km and hence they do not 
decay. However for $\dm \geq 10^{-3}$ eV$^2$    
and large mixing angles there will be substantial $\nu_e - \nu_x$ conversion  
in conflict with the SK atmospheric neutrino and CHOOZ data \cite{SK,CHOOZ}. 

In conclusion, neutrino decay solution to the solar neutrino problem is
ruled out at 99.96\% C.L. from the current 
data on total rates. For the SK spectrum data however, one can get much
better fits, for $\nu_2$ lifetimes consistent with cosmological and supernova
constraints. 
Although the best-fit for the spectrum data comes in the region 
where the probability is energy dependent, if $X_{B}$ is allowed 
to vary the $\chi^2$ becomes flat over the entire range of 
parameter space. This implies that even energy independent 
suppression of the $^8B$ flux is allowed. Even if $X_{B}$ is 
fixed at 1, the decay scenario can give 
$\chi^2$ comparable to the best-fit in the energy 
independent high $\alpha$ regime.  
For the $^{8}{B}$ flux
normalisation factor a wide range  0.4 $< X_{B} < 7.2$ is allowed at 
90\% C.L. just from the SK spectral data, if neutrino decay is operative. 
This is much broader than the range allowed by SSM uncertainties.  
For the rate+spectrum analysis the decay solution is disfavoured 
at more than 98\% C.L. even if $X_{B}$ is allowed to take arbitrary 
values.  
 
\vskip 10mm
\noindent
{\small The authors thank A. Raychaudhuri for his involvement 
during the development of the solar neutrino code that has been 
used in this paper. 
They also wish to thank S. Pakvasa and K. Kar for             
discussions and the organisers of WHEPP-6 where this 
work was initiated. D.M. acknowledges financial support from the 
Eastern Center of Reasearch in Astrophysics, India.}

\newpage
\begin{center}
Figure Captions
\end{center}

\noindent
Fig. 1: The variation of the survival probability $P_{\nue\nue}$ (thick lines) 
and the exponential decay term $\exp(-\alpha$ L/E) (thin lines) with $\alpha$, 
for two different values of neutrino energies. The solid curves correspond 
to neutrino energy = 7 MeV while the dotted curves are for 
neutrino energy = 13 MeV. For both the cases $\sin^2\theta$ is fixed at 
0.51.
\vskip 10mm

\noindent
Fig. 2: The variation of $X_{B}$ with $\alpha$ for three different 
values of $\sin^2\theta$ shown in the plot. 
Each point on these curves is obtained 
by keeping $\alpha$ and $\sin^2 \theta$ fixed and determining the 
$X_{B}$ corresponding to the minimum $\chi^2$. 

\vskip 10mm
\noindent
Fig. 3: The variation of $\chi^2$ with 
(a) $\alpha$ for $X_{B}$ unconstrained (solid line) and fixed at 1.0 
(dashed line), while $\sin^2 \theta$ is kept unconstrained for both curves; 
(b) $\sin^2 \theta$ keeping both $\alpha$ and $X_{B}$ unconstrained for  
the solid curve and with $\alpha$ unconstrained 
and $X_{B}$ fixed at 1.0 for the dashed curve;
(c) $X_{B}$ keeping both $\alpha$ and $\sin^2 \theta$ unconstrained. 
The dotted line shows the 90\% C.L. limit for 3 parameters 
($\chi^2 = \chi^2_{min} + 6.25$) and the dash-dotted line gives the 
corresponding limit for 2 parameters 
($\chi^2 = \chi^2_{min} + 4.61$).
\begin{figure}[p]
\epsfxsize 16 cm
\epsfysize 17 cm
\epsfbox[25 151 585 704]{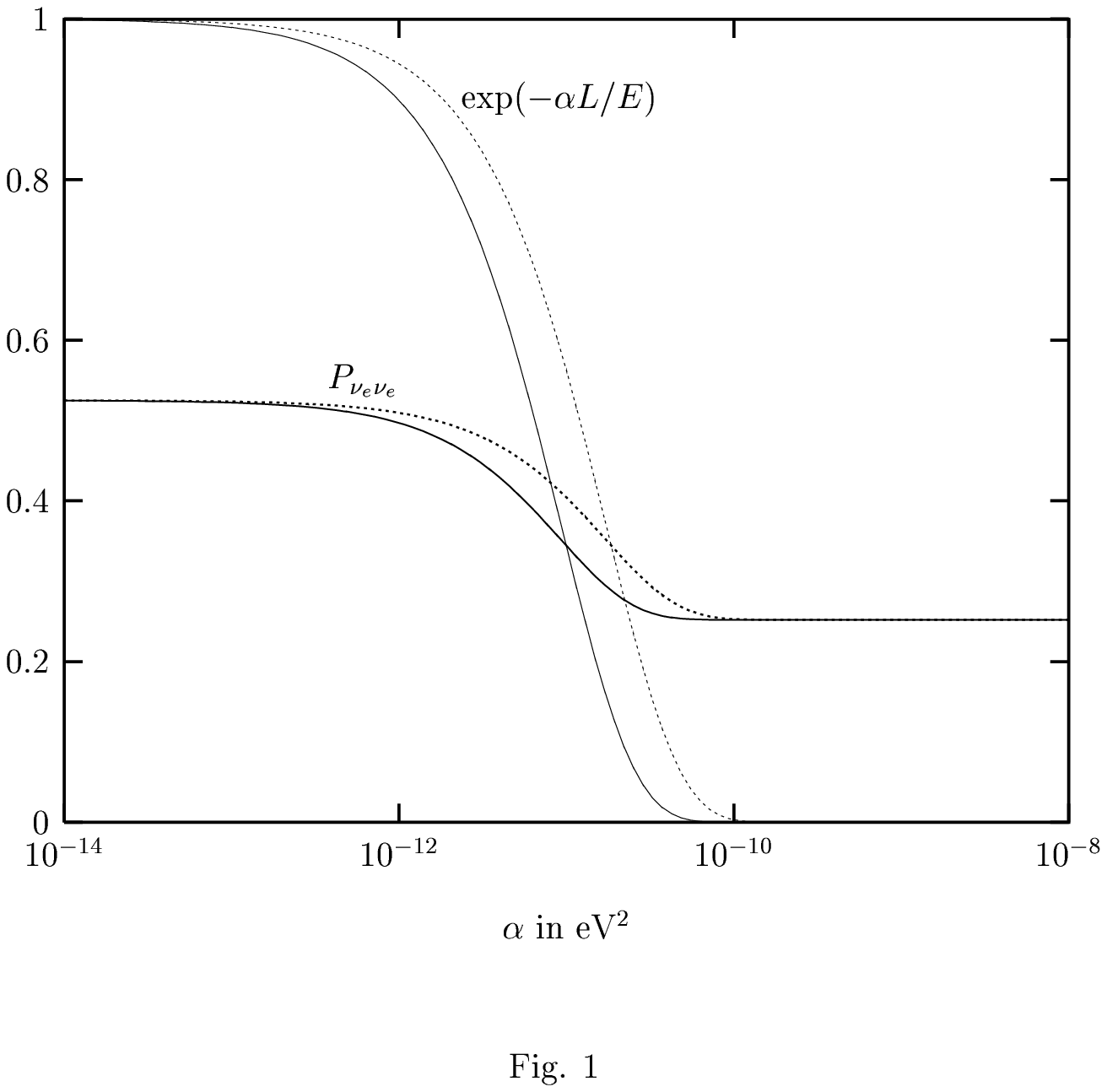}
\end{figure}

\begin{figure}[p]
\epsfxsize 16 cm
\epsfysize 17 cm
\epsfbox[25 151 585 704]{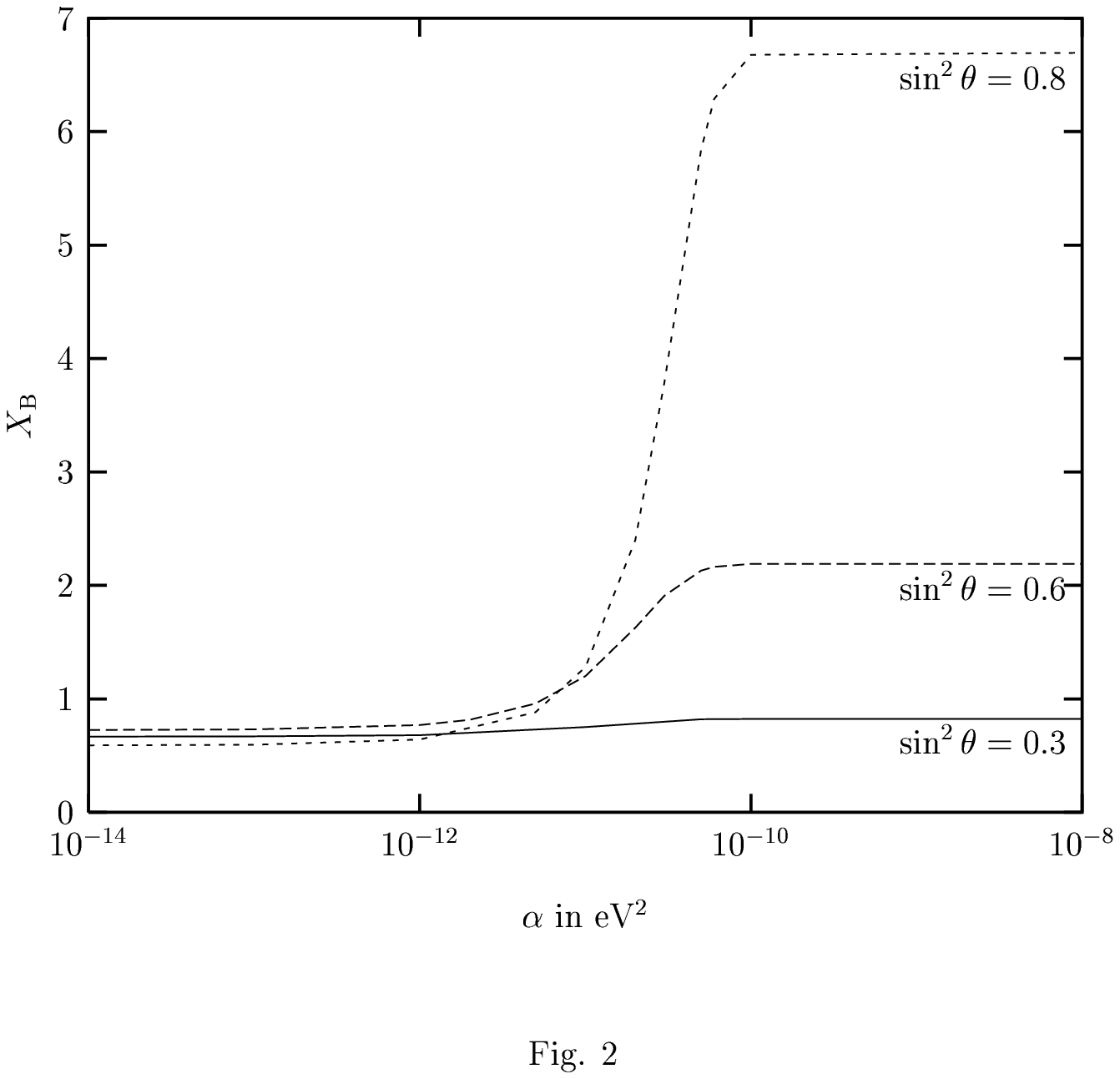}
\end{figure}

\begin{figure}[p]
\epsfxsize 16 cm
\epsfysize 17 cm
\epsfbox[25 151 585 704]{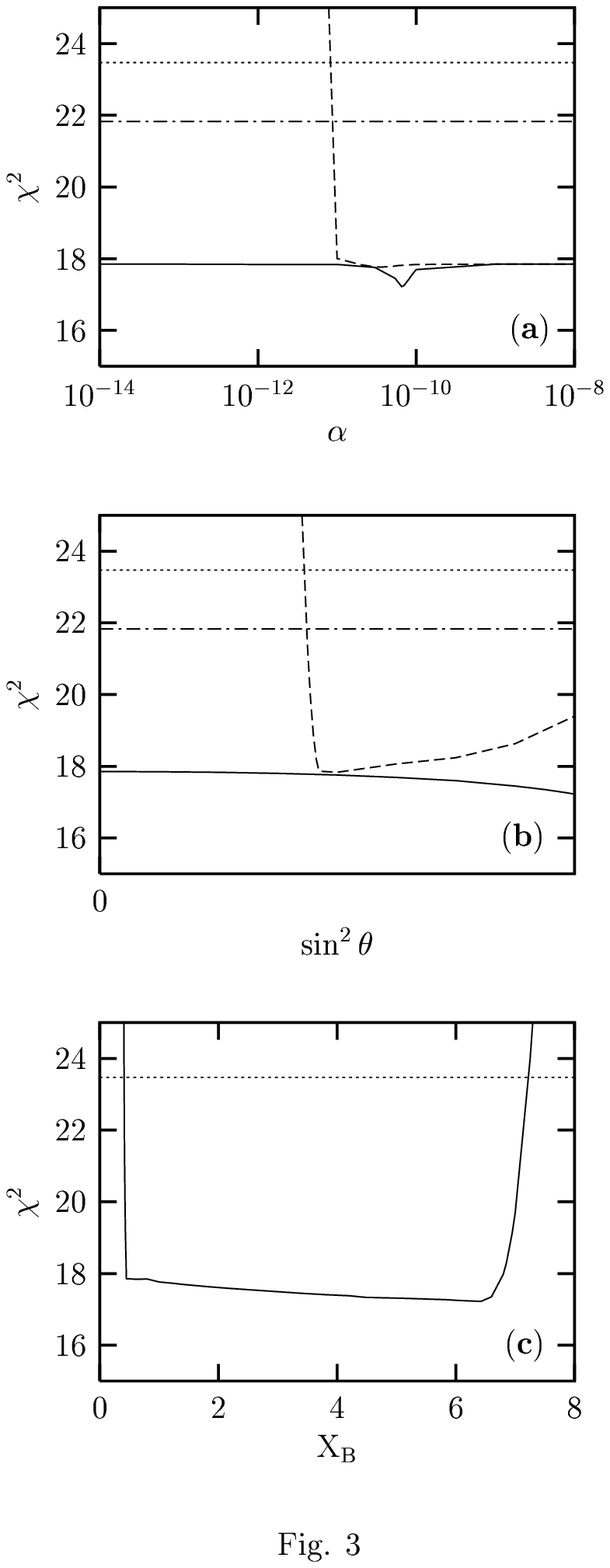}
\end{figure}

\end{document}